\documentclass[11pt,letter]{article}
\usepackage[pdftex]{graphicx}
\usepackage[numbers,sort&compress]{natbib}
\usepackage{multirow}
\usepackage{amssymb}
\usepackage[outercaption]{sidecap} 
\usepackage{amsmath,amsfonts}
\usepackage{url}
\usepackage[T1]{fontenc}
\usepackage{upquote}
\usepackage{longtable}

\usepackage{url}
\usepackage[x11names]{xcolor}
\usepackage{footmisc}
\usepackage{enumitem}

\begin{document}
\parskip 1ex
\pagestyle{empty}

%
\newcommand{\D}{\ensuremath\mathrm{d}}
\begin{center}
{\large \bf Towards a Common Format for Computational Material Science Data}

{Luca M. Ghiringhelli$^{1,}$\footnote{ghiringhelli@fhi-berlin.mpg.de},
Christian Carbogno$^{1}$, 
Sergey Levchenko$^{1}$, 
Fawzi Mohamed$^{1}$, 
Georg Huhs$^{2,3}$, 
Martin L\"{u}ders$^{4}$, 
Micael Oliveira$^{5,6,}$\footnote{mjt.oliveira@ulg.ac.be},
and Matthias Scheffler$^{1,7}$
}

{\small $^1$Fritz Haber Institute of the Max Planck Society, Berlin, Germany}\\
{\small $^2$Barcelona Supercomputing Center, Spain}
{\small $^3$Humboldt-Universit\"{a}t zu Berlin}\\
{\small $^4$Daresbury Laboratory, UK}
{\small $^5$University of Li\`{e}ge, Belgium}\\
{\small $^6$Max Planck Institute for the Structure and Dynamics of Matter, Hamburg, Germany}\\
{\small $^7$Department of Chemistry and Biochemistry and Materials Department, University of California--Santa Barbara, Santa Barbara, CA 93106-5050, USA}\\
\end{center}

\begin{abstract}

Information and data exchange is an important aspect of scientific progress. In computational materials science, a prerequisite for smooth data exchange is standardization, which means using agreed conventions for, e.g., units, zero base lines, and file formats. There are two main strategies to achieve this goal. 
One accepts the heterogeneous nature of the community which comprises scientists from physics, chemistry, bio-physics, and materials science, by complying with the diverse ecosystem of computer codes and thus develops ``converters'' for the input and output files of all important codes. These converters then translate the data of all important codes into a standardized, code-independent format.
The other strategy is to provide standardized open libraries that code developers can adopt for shaping their inputs, outputs, and restart files, directly into the same code-independent format. We like to emphasize in this paper that these two strategies can and should be regarded as complementary, if not even synergetic.\\
The main concepts and software developments of both strategies are very much identical, and, obviously, both approaches should give the same final result. 
In this paper, we present the appropriate format and conventions that were agreed upon by two teams, the Electronic Structure Library (ESL) of CECAM and the NOMAD (NOvel MAterials Discovery) Laboratory, a European Centre of Excellence (CoE). This discussion includes also the definition of hierarchical metadata describing state-of-the-art electronic-structure calculations.
\end{abstract}

\newpage

\section{Introduction}
To aid and guide the search for new and improved materials, computational materials science is increasingly employing 
``high-throughput screening'' calculations \cite{Ceder09,Wood09,Wolverton10,Ceder11,Rajan11,Jakobsen12,Curtarolo13}.
In practice, this means that computational material scientists produce a huge amount of materials data on their local workstations, computer clusters, and supercomputers using a variety of very different computer programs, in this domain usually called ``codes''. Though being extremely valuable, this information is hardly available to the community, since most of the data is stored locally or even deleted right away. In publications, typically, only  a small subset of the calculated results is reported, namely that which is directly relevant for the very specific topic addressed in the actual manuscript. Although several repositories have been created and maintained in the past for domain-specific applications, these typically do not store the full inputs and outputs of all calculations. And, typically, they are restricted to results obtained by just one or very few codes. Since 2014, this situation is changing: the NOMAD Repository (\texttt{https://NOMAD-Repository.eu}) \footnote{The NOMAD Repository is a joint activity of the groups of Matthias Scheffler (FHI), Claudia Draxl (HU) and the Max Planck Computing and Data Facility: \texttt{https://NOMAD-Repository.eu}. Data is stored for at least 10 years and a DOI is provided so that data is citable. \texttt{http://NOMAD-repository.eu} is the only repository for materials science accepted by Nature Scientific Data (\texttt{http://www.nature.com/sdata/})} is designed to meet the increasing demand for storing scientific data and making them available to the community. It is a unique facility, as it accepts (and requests) input and output files of all important computer codes used in computational materials science, and it guarantees to store these data for at least 10 years.\\ 
Currently, the repository holds the information of over 3 million calculations which corresponds to more than a billion CPU-core hours. While this has been and is useful for its purpose of data sharing via a repository (enabling the confirmatory analysis of materials data, their reuse, and repurposing), the data is very heterogeneous, just as they are provided by the different codes.
Thus, they are not useful for data analytics and extensive comparisons, because the input and output files of different codes typically use different units, representations, and file formats. \\
Advancing the sharing and comparison of such data is a pressing issue that needs to be addressed, as exemplified in recent initiatives. 
For instance, the NOMAD Laboratory CoE \cite{NOMAD} is presently building a unified, code-independent database, so that big-data analytics techniques can be applied to obtain unprecedented insight from the vast amount of already existing calculations. \\
In a similar spirit, the E-CAM CoE \cite{ECAM}, which was recently established by CECAM to build an e-infrastructure for software, training, and consultancy in simulation and modeling, is committed to actively support the development and adoption of software libraries and standards within the electronic structure community. One measure aiming at this is CECAM's Electronic Structure Library (ESL) initiative \cite{ESL}, which includes in its plans the development of an Electronic Structure Common Data Format (ESCDF). The ESCDF provides a standardized data format and an Application Programming Interface (API) that every code can use.
Code interoperability is also strongly driven by communities, using the results for further analysis, such as spectroscopy calculations, based on converged and relaxed ground state results. These applications have been at the heart of the European Theoretical Spectroscopy Facility (ETSF) \cite{Gonze08} activities, and are now a central part of the efforts of the COST network EUSpec \cite{EUSPEC}.\\
As mentioned above, the $\Psi_k$ community is using many different computer codes. 
A list of the most important electronic-structure codes in the materials science community is given in Table 1. The codes are ranked by the number of citations of the last five years and their main features are summarized.

\begin{table}[b!]
\footnotesize
\begin{tabular}{l|l|r|l|l|l}
Code & Search Name & Citations & Description & License\tabularnewline
 & & (2011-15) & & \tabularnewline
Gaussian & Frisch \cite{notetable}& 18\,700 & AE, PSP, GTO & C\tabularnewline
VASP & Kresse  & 16\,500 & PSP, PW, PAW & C(O)$^\textrm{(a)}$\tabularnewline
CASTEP & Payne & 5\,970 & PSP, PW, PAW & C(O)$^\textrm{(b)}$\tabularnewline
GAMESS & Gordon & 5\,360 & AE, PSP, GTO & F\tabularnewline
WIEN2k & Blaha & 4\,880 & AE, (L)APW & C\tabularnewline
Quantum ESPRESSO &  Giannozzi \cite{notetable} & 4\,810 & PSP, PW & G\tabularnewline
Molpro & Werner & 4\,190 & AE, PSP, GTO & C\tabularnewline
SIESTA & Soler & 4\,040 & PSP, NAO & G\tabularnewline
TURBOMOLE &  Ahlrichs & 3\,730 & AE, PSP, GTO & C\tabularnewline
ADF &  Baerends & 2\,860 & AE, STO & C\tabularnewline
DMol$^3$ & Delley \cite{notetable}& 2\,893 & AE, PSP, NAO & C\tabularnewline
ORCA & Neese & 2\,540 & AE, PSP, GTO & F\tabularnewline
CRYSTAL & Dovesi \cite{notetable}& 2\,225 & AE, PSP, GTO & C$^\textrm{(b,c)}$\tabularnewline 
MOPAC & Stewart & 2\,070 & PSP, STO & F(O)$^\textrm{(d)}$\tabularnewline
ABINIT & Gonze & 1\,870 & PSP, PW, PAW, WLT & G\tabularnewline
Q-Chem & Shao \cite{notetable} & 1\,740& AE, PSP, GTO & C\tabularnewline
Jaguar & {\em Schr\"{o}dinger} & 1\,670 & AE, PSP, GTO, STO & C\tabularnewline
Dalton & \AA{}gren & 1\,580 & PSP, GTO & F\tabularnewline
NWChem & Valiev & 1\,410 & AE, PSP, GTO, PW, PAW & G\tabularnewline
MOLCAS & Lindh & 1\,200 & AE, PSP, GTO & C$^\textrm{(e)}$\tabularnewline
CP2K & Hutter \cite{notetable}& 1\,190 & AE, PSP, GTO, PW & G\tabularnewline
ACES III & Bartlett \cite{notetable} & 1\,150 & PSP, GTO & G\tabularnewline
CPMD & Hutter & 938 & PW, PSP & F\tabularnewline
TB-LMTO-ASA & Andersen \cite{notetable}& 840 & AE, LMTO, $N$MTO & G\tabularnewline
octopus & Rubio & 792 & PSP, RS & G\tabularnewline
CFOUR & Stanton \cite{notetable}& 738 & AE, PSP, GTO & F\tabularnewline
GPAW & Mortensen & 624 & PSP, PW, PAW, RS, NAO & G\tabularnewline
DIRAC & Saue \cite{notetable}& 579 & AE, LAO & F(O)\tabularnewline
CASINO &  Needs \cite{notetable}& 475 & AE, PSP, PW, GTO, STO, NAO & F$^\textrm{(d)}$\tabularnewline
FPLO & Koepernik & 459 & AE, NAO & C\tabularnewline
FHI-aims & Blum & 407 & AE, NAO & C(O)\tabularnewline
OpenMX & Ozaki  & 359 & PSP, NAO & G\tabularnewline
COLUMBUS & Lischka & 333 & AE, GTO & F\tabularnewline
Smeagol & Lambert & 302 & PSP, NAO & F\tabularnewline
Elk & Dewhurst \cite{notetable}& 225 & AE, LAPW & G\tabularnewline
\end{tabular}
\end{table}

\begin{table}[t!]
\footnotesize
\begin{tabular}{l|l|r|l|l|l}
Code & Name & Citations & Description & License\tabularnewline
 & & (2011-15) & & \tabularnewline
Yambo & Marini & 220 & PSP, PW & G\tabularnewline
FLEUR & Bl\"{u}gel  & 212 & AE, LAPW & F(O)\tabularnewline
ONETEP & Skylaris & 204 & PSP, PW/RS, PAW, NAO & C(O)\tabularnewline
CONQUEST O(N) & Bowler \cite{notetable}& 201 & AE, PSP, NAO, PW & I\tabularnewline
Psi4 &  Sherrill & 187 & AE, GTO & F(O)\tabularnewline
TeraChem & Martinez \cite{notetable}& 175 & AE, PSP, GTO & C\tabularnewline
exciting & Draxl \cite{notetable}& 171 & AE, LAPW & G\tabularnewline
BigDFT & Genovese & 141 & PSP, WLT & G\tabularnewline
BerkeleyGW & Louie & 140 & AE, PSP, PW, NAO, RS & F(O)\tabularnewline
PARSEC & Saad & 135 & PSP, RS & G\tabularnewline
ATOMPAW & Holzwarth & 105 & AE, PSP, PW, PAW & F(O)\tabularnewline
\end{tabular}
\caption{\footnotesize List of the electronic-structure codes with more than 100 citations in the 2011-2015 period. The number of citations is  determined via Google Scholar, with a search performed in April 2016, by searching for the name of the code (as reported in the first column) {\em and} the name of one of the main developers (or {\em company} that develops and commercializes the product, as reported in the second column). The reason of the second search criterion is that the name of several codes have different meanings. We found that the combination of the two column gave very few, if any, false positive results. Nonetheless, in some cases we had to further adjust the search criterion to eliminate the false positive results; the actual search strings are listed in note \cite{notetable}. The significance of the citation numbers should not be overrated (as in any citation analysis). For example, young codes that were only developed in the last 7-8 years may have a high gradient of their employment in the community but still have a rather low citation index in the table. Nevertheless, we believe that the general impression provided by the table is correct. The fourth column lists the main features of each code and the fifth the type of license. A list including also force-field based codes is maintained at: \texttt{https://www.nomad-coe.eu/index.php?page=codes}. The meaning of the abbreviations is the following. For ``Description'', AE: All Electron, PW: Plane Wave, GTO: Gaussian-Type atomic Orbitals, NAO: Numeric Atomic Orbitals, STO: Slater-Type Orbitals, LAPW: Linearized Augmented Plane Wave, PAW: Projector Augmented wave, RS: Real Space, PSP: pseudopotential (including also ECP, effective core potential), LAO: London Atomic Orbitals, LMTO/$N$MTO: (Linear) Muffin-Tin Orbitals / $N^\textrm{th}$-order MTO, WLT: Wavelet; for ``License'', G: GPL or LGPL (includes also GPL compatible licenses, such as Educational Community License or Apache 2), F: Free (other licenses), C: Commercial / charged (usually, with a smaller cost for academics compared to non-academics), (O): Open source, I: Individual-basis license (via contacting the authors), $^\textrm{(a)}$: Free for academic use in Austria, $^\textrm{(b)}$: Free for academic use in the UK, $^\textrm{(c)}$: Free demo serial version (max 20 atoms/cell), $^\textrm{(d)}$: Commercial for non-academic use, $^\textrm{(e)}$: version 8.2 will have license G.}
\end{table}

In this highlight paper, we will discuss the challenges of electronic-structure codes, but note that the NOMAD Laboratory CoE also includes force-field based codes. The various 
electronic-structure codes differ by many conceptual and numerical aspects, first of all by the type of basis sets that are employed. 
Besides, some codes describe all electrons, others use a simplified description of the core electrons. On one side, this results in very different data representations; on the other side, it is not trivial to compare certain quantities like energies and wavefunctions.\\
In the remainder of this paper, we will define a common code-independent representation for all relevant quantities (e.g., structure, energy, electronic wave functions, trajectories of the atoms, etc.).
The ideas and concepts presented here are also the result of discussions that took place at a CECAM/Psi-k workshop, held in Lausanne in January 2016, that was attended by experts and key developers of more than 20 of the most important electronic structure codes. 

\section{The conversion layer}
In this section, we discuss the key issues for converting the information present in the inputs and outputs of electronic-structure codes into a common format. More specifically, this discussion addresses: i) The metadata infrastructure for storage and retrieval of the code-independent quantities; ii) the uniform file format as defined by the ESCDF team; iii) the zero-level reference for energy-related quantities; iv) the representation for the electronic and vibrational band structures and density of states; v) the compact representation of scalar fields such as wave functions, electron densities, and exchange-correlation potentials; vi) the unified representation of quantities related to excited-state calculations ($GW$, Bethe-Salpeter). 
Furthermore, we discuss the general challenge, touching all the above points, of establishing error bars and confidence levels, with respect to the adopted numerical settings, for each stored/retrieved calculation.

\subsection{Metadata for the code-independent format}
Metadata is the name or label that characterizes corresponding values.
For example ``XC\_method'' is a metadata name and ``LDA'' may be the corresponding value. Thus, if one thinks of storing data as {\em key}--{\em value} pairs (as in a dictionary), the {\em key} is the metadata. \\
There are several examples of metadata approaches in computational materials science, the most prominent being ChemML, the Chemical Markup Language \cite{CML}, CIF, the Crystallography Information File \cite{CIF}, code specific implementations such as the VASP \cite{VASP} or Molpro \cite{MOLPRO} XML outputs, or very simple and intuitive file formats like XYZ \cite{XYZ} for the storage of the configuration of a system (atomic species, coordinates, also allowing for comments).\\
In general, a metadata structure is built {\em a priori}, by starting from a list of names that identify the needed concepts and quantities. Code outputs and file formats are then designed to reflect the {\em a priori} metadata structure. In principle, {\em a priori} metadata approaches aim to be as exhaustive as possible; in practice, they are typically designed with a specific scientific field, application and/or code in mind, as the examples above show. Conversely, the ``NOMAD Meta Info'' \cite{metainfo}, i.e., the metadata defined and used for the NOMAD code-independent data format, is defined {\em a posteriori}, by starting from the existing inputs and outputs of many different codes stemming from different scientific fields and thus with quite diverse feature sets.
From a computer science point of view, the {\em a posteriori} approach is much more challenging than the {\em a priori} one, since it 
requires a more flexible, extensive, and generalized infrastructure~\cite{metaweb}:  On the one hand, the diverse quantities stored by each code need to find the proper place in the global hierarchy. On the other hand, not all codes' inputs and outputs contain all the defined metadata; therefore, having unassigned metadata when sorting a given input/output in the code-independent format must not create problems. From a physics, chemistry, and materials science point of view, the  {\em a posteriori} approach has the critical advantage that essentially all the information contained in any input/output file of any considered code is recognized and stored into the properly identified {\em key}--{\em value} pair. Intrinsically, this guarantees a truly exhaustive coverage of all properties, even those that currently might not appear of interest. In turn, this lays the founding for authentic big data mining approaches that unveil hitherto unexpected correlation and relationships.\\
In NOMAD Meta Info \cite{metainfo}, the {\em key} is not a simple string but a more structured object, with several attributes. The {\em name} is an attribute, a string that must be unique, well defined, intuitive and as short as possible\footnote{As the {\em names} will be the string frequently present in parser scripts when reading inputs/outputs and assigning a value to a given metadata, as well as used in query scripts in order to locate a particular metadata, it is convenient to keep these strings as short as possible.}. It is used to identify the metadata and therefore used to associate {\em values} with their metadata.
The {\em description} is another attribute and contains an extensive human-readable text that clarifies the meaning of the metadata. \\
Another important attribute of a metadata is its {\em type}. Concrete {\em values}, scalars, strings, or (multidimensional) arrays,  that are extracted by the parsers that read input and output files, have an associated  {\em concrete-type} metadata.
These {\em values} are organized in groups to which {\em section-type} metadata are associated. In computational informatics, these {\em sections} would correspond to tables of a relational database model \footnote{In a relational database, data are organized in tables, where rows represent instances of some entity (e.g., a customer, a product) and the columns values attributed to that instance (e.g., the address, the price). Rows are identified by unique keys.}, where the {\em values} would be the rows, and the {\em concrete} metadata the columns.\\
By considering the different steps that define an electronic-structure calculation run (i.e., from the invocation of the code to the completion of the task described in the input files, or the interruption due to several reasons), the following main {\em section-type} metadata are defined:
\begin{itemize}[noitemsep,nolistsep]
 \item ``section\_run'': represents a single run of the program,
 \item ``section\_method'': contains the information defining the theory level and convergence parameters,
 \item ``section\_system'': contains the specifics of the system configuration (atom species, coordinates, lattice vectors), 
 \item ``section\_single\_configuration\_calculation'': contains the results for a physical system as defined in a single ``section\_method'' and a single ``section\_system'',
 \item ``section\_scf\_iteration'': contains the results of a single self-consistency iteration.
\end{itemize}
The chosen hierarchy reflects that {\em sections} can be nested, meaning that each outer one can contain one or more inner {\em sections},~e.g.,~each ``section\_single\_configuration\_calculation'' typically contains multiple iterations of ``section\_scf\_iteration''. Also, the different {\em sections} depend on each other,~e.g.,~the outcome in ``section\_single\_configuration\_calculation'' depends on the system, method, and program defined in the higher layers of the structure. 

\begin{figure}[h!]
  \begin{minipage}[c]{0.70\textwidth}
    \includegraphics[width=0.9\textwidth,clip]{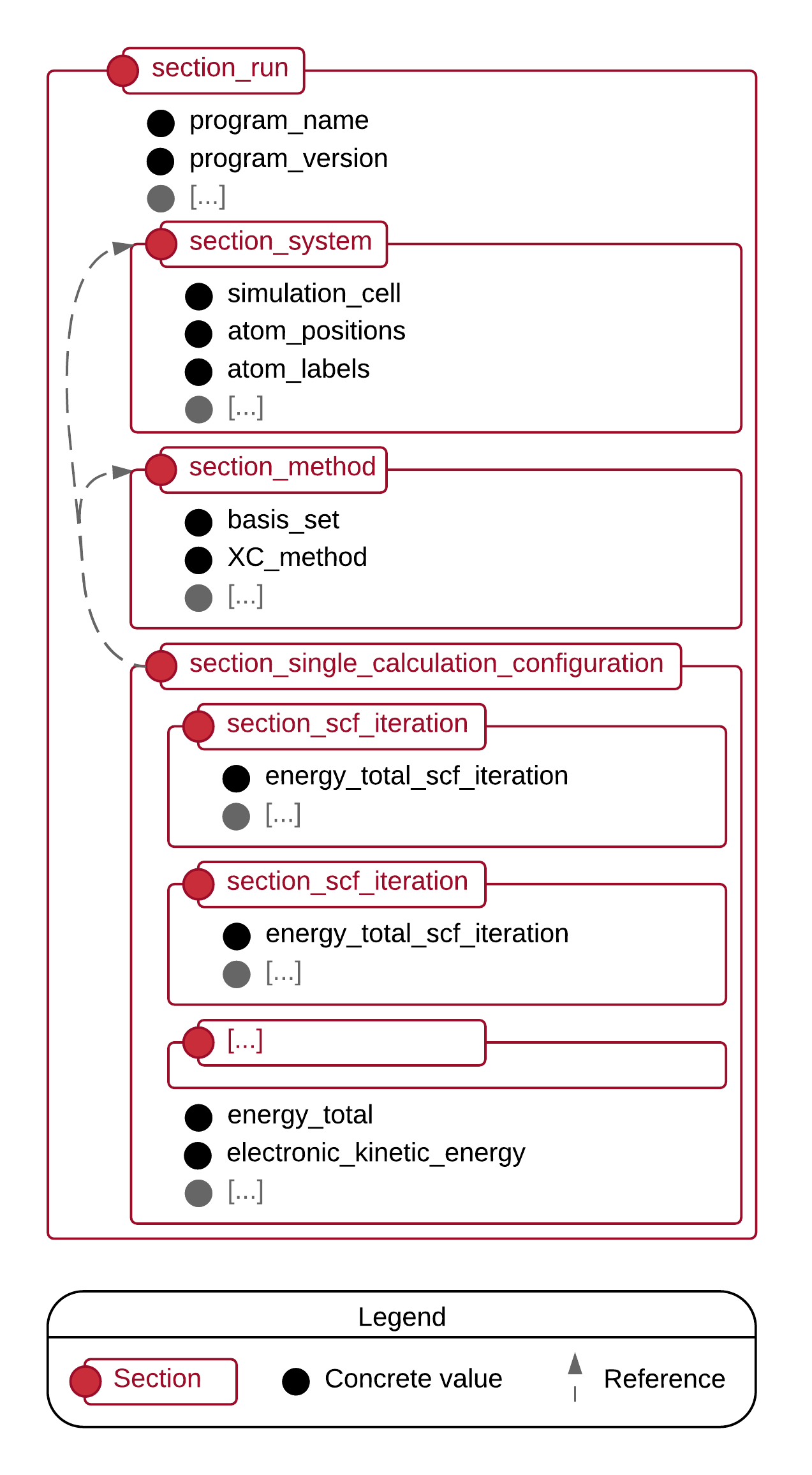}
  \end{minipage}\hfill
  \begin{minipage}[c]{0.30\textwidth}
    \caption{
      A simplified metadata structure, according to NOMAD Meta Info \cite{metainfo}, for a simple electronic-structure calculation is shown.
      The metadata indicated in black are of {\em concrete} type. The {\em names} are self explaining (the full {\em description} is given in \cite{metaweb}). The metadata indicated in red are of type {\em section} and the content of the {\em section} defined by them is inside the red box. {\em Sections} can also have references, indicated by the dashed arrows, to other {\em sections} besides the parent {\em section}. 
      The layout of the metadata in the figure is reminiscent of the JSON file, that is one of the file formats in which they are effectively stored. } \label{metafig}
  \end{minipage}
\end{figure}

The power and flexibility of this approach is schematically displayed in Fig. \ref{metafig}: We show the (simplified) representation following the metadata structure (according to NOMAD Meta Info \cite{metainfo}) of a simple calculation (a ``section\_run'') where one structure (defined in ``section\_system'') is evaluated with one electronic-structure method (defined in ``section\_method''). The results are reported in a single ``section\_single\_configuration\_calculation'', that in this example contains, besides the final results, also results from various scf iterations (``section\_scf\_iteration''). The metadata of {\em concrete} type (black font) have also {\em units} (in NOMAD Meta Info, we always use SI units) as attribute. In practice, ``energy\_total'' could have {\em value} ``$-1.344 \cdot 10^{-20}$'', with {\em units} ``J'', for joule. {\em Concrete} metadata with associated values are contained in {\em sections} (red font) and {\em sections} can also be contained in parent {\em sections}, as graphically indicated by the boxes. {\em Sections} can also have references (dashed arrows) to other {\em sections}. In the present case, this is needed to relate the results contained in ``section\_single\_configuration\_calculation'' with the physical model used to obtained them (``section\_method'') and the geometrical structure used as input (``section\_system''). In more complex cases, several ``section\_single\_configuration\_calculation'' can have references to the same ``section\_method'' (e.g., a geometry optimization) and/or to the same ``section\_system'' (e.g. the same system calculated with a self-consistent electronic-structure method which is used as a starting point for a many-body perturbation method), see \cite{metadata}. 
The standard definition of NOMAD Meta Info is maintained in a git repository \cite{metaweb} and contributions are welcome. The current metadata structure can be browsed at \cite{metainfo}.\\
%
NOMAD Meta Info \cite{metainfo} is kept independent of the actual storage format as much as possible; it is a conceptual model, and in principle not bound to any specific storage method. 
In previous efforts, the metadata structure was normally seen as keys without any structure, consequently the relationships between various key--value pairs was expressed using the means of the specific storage format that was chosen.
To actually make the data usable, we do express the structure of the values in the metadata using sections and references, and define the types used to store a concrete value.
We knowingly limited the allowed types and the kind of structuring we support, to strike a balance between being able to represent any quantity in a logical way, and being able to use this representation in multiple places and formats.
Some things might be represented slightly better for example in JSON, XML, HDF5, CIF,... using some special features of each specific format, but then it is difficult to transfer those features in a natural way to another representation, especially if one does not know automatically which structuring of a specific format should be preserved, and what can be ignored with little loss.\\
We feel that a simplified view supporting multiple representations allows one to choose different formats depending on what needs to be optimized in a specific application, and simplifies what the user has to understand.
We officially support two formats to store data using the NOMAD Meta Info: the human readable JSON and HDF5, a binary format that can store efficiently large arrays and higher dimensional objects.
This covers already quite different trade-offs with respect to human readability and storage efficiency, and has influenced and validated the choices done in the NOMAD Meta Info.
The data could also be represented in other formats (CIF, XML, Relational Databases,...), and an official mapping to some of them could be added in the future.\\
More details, including details on the practical implementation are given in Refs. \cite{metaweb} and \cite{metadata}. 

\subsection{The electronic structure common data format (ESCDF).} 
Whenever code developers are willing to design their inputs and outputs in a standardized format, the ESCDF library will 
provide the necessary common ground.\\
In particular, the ESCDF is focused on giving tools for developers to save the data produced by their code for the 
purpose of restarting a calculation, sharing the data with other codes, or further processing.
This is achieved by standardizing the quantities to be put in an output file and via adopting self-describing formats 
like HDF5 or NetCDF, which are extendable and allow the inclusion of metadata needed to interpret them. 
The synergy between the NOMAD Laboratory CoE and ESCDF lies in the fact that, in the conversion layer, the NOMAD Laboratory will use
the ESCDF file format, while the ESCDF will use the NOMAD metadata.
For a given code to be able to use the information produced by another code, the mere capability of reading it from a file is not sufficient. 
Indeed, as explained in the Introduction, the various electronic-structure codes use very different data representations, 
of which the type of basis set is the most paradigmatic example. This means that, after reading the data, it might be necessary to 
perform some conversions. Some are quite straightforward, like changing units, but other conversions can be quite involved,
either requiring complex algorithms and/or be computationally demanding. For example, to be able to use wavefunctions that were written 
as a linear combination of Gaussian-type atomic orbitals, a plane wave code would have to perform a change of basis. 
Such transformations are beyond the scope of the ESCDF, but are an essential component of the NOMAD converters.
From here the synergies between the two projects become clear: the ESCDF provides the tools for a standardized access 
to the data stored within the files, while the NOMAD conversion layer converts the data to a code-independent representation.\\
The first version of the ESCDF will include specifications to read/write the following type of data: geometry/structure 
of the system, basis sets, densities, potentials, and wavefunctions. At this point, the specifications do not aim at covering
exhaustively all the quantities that an electronic code might need to read/write. Instead, the focus is on making sure the specifications
are flexible and extendable. Because of its hierarchical structure and greater 
flexibility, HDF5 was chosen over NetCDF as the underlying file format. Each type of data is stored in an HDF5 group, which
can be arranged in a way that is similar to a file system. This allows to store data for different use cases. For example,
the most common use case is performing one calculation for one system, but the format should also be able deal with several
systems that are calculated simultaneously. Using the NOMAD metadata and the experience accumulated by a previous standardization
effort led by the European Theoretical Spectroscopy Facility (ETSF)~\cite{Gonze08}, a set of specifications have already been agreed for the 
geometry/structure and work is currently underway regarding the basis sets and scalar fields.\\
The associated software library and corresponding API will focus on flexibility, extensibility, and performance in order to maximize its usefulness and adoption by the 
community of code developers. In particular, we aim at providing an API that does not force code developers to change the way
how they store their data in memory, even in the case of parallel applications where the data is distributed among different processors.
This is technically challenging, but essential if one wants to allow the code developers to focus on implementing new features and exploring new ideas
instead of spending time porting and optimizing their codes to specific computer architectures.

\subsection{A common energy zero for total energies.} 
Many physical properties, e.g. forces, elastic constants, energy barriers, depend on total-energy differences. Therefore, they are well defined and the comparison of results from different codes is readily possible. 
To make also total energies stemming from different codes comparable, it is necessary to define a reference energy scale. To achieve this goal, a simple, pragmatic computational prescription viable for all codes is necessary.\\
Our idea is to define relative energies, where the energy of conveniently defined reference atoms is subtracted from the total energies calculated by each code. Ideally, the reference atoms would be isolated neutral atoms and we would then have formation/atomization energies as code-independent energies. However, 
it is well known that calculations for isolated atoms can be problematic for solid-state electronic-structure codes that use plane-wave basis sets (including augmented planewave ((L)APW) codes and alike). These codes are designed to study periodic systems, and the description of isolated atoms then requires large unit cells to ensure that the atoms do not interact. This makes the calculations expensive, and it may even prevent a systematic convergence study of all numerical settings, such as basis sets and grids, etc., for some atoms. 
To bridge the gap between periodic and non-periodic codes, both free atoms and simple bulk systems should be used as reference systems. \\
The coexistence of several reference-energy definitions is not a limit in the comparison as long as at least one code can encompass all definitions and evaluate all the reference values. Such code (or group of codes), would serve as a ``Rosetta stone'': This  2200 years old stele
enabled the comparison and identification of Ancient Egyptian hieroglyphs, Demotic script, and Ancient Greek. 
Here is a brief summary of the adopted strategy, which is thoroughly discussed in Ref. \cite{rosetta}.\\
When using free atoms as reference, the simplest choice is to define fully converged atoms (with respect to basis sets and integration grids). Their energies are calculated once for each physical model (exchange-correlation -- xc -- treatment) and, for pseudopotential-based codes, for each pseudopotential. In practice, free atoms are evaluated as spin unpolarized non-relativistic, in order to allow for a safe comparison among different codes where the implementation of the spin and relativistic treatments may be very different. However, other sets with spin-polarized atoms and selected relativistic treatments could be also stored, in order to obtain (atomization) energies that are closer to a physical meaning. 
Atomic energies evaluated at the same numerical (besides the physical) settings as the calculation of interest can also be considered. On one side this strategy allows for a well known partial cancellation of numerical errors, on the other side the number of stored entries can potentially grow uncontrolled. To avoid this, a machine learning strategy that predicts the best atomic energy to be subtracted on the basis of a minimal, informative amount of stored information could be designed.\\
When using period bulk systems as reference, in a so-called ``thermodynamic approach'', the practical choice is to use the same crystal structures as used in K. Lejaeghere \emph{et al.} \cite{Cottenier, Cottenier2}. The few gaps in these publications (lanthanides and actinides) can be filled by using the structures reported in the CRC Handbook \cite{CRC}. For species like O and N that form a molecular solid, where the combination of covalent and weak interaction may add numerical noise, one could also consider to use a binary compound where the other species of the binary compound forms a covalent or metallic crystal. Also in this case the practical choice is to use {\em fully converged} (with respect to basis sets, integration grids, and $k$ grids) reference calculations, one for each physical model (including pseudopotential). Considering to subtract reference solids at the same numerical settings as the calculation of interest implies the same pros and cons as for the free atoms.

\subsection{Electronic and vibrational properties of solids}
Electronic band structures are typically represented along high-symmetry paths in the first Brillouin zone. 
In literature one can find a large variety of such paths, differing in directions and sequence of the path-segments. 
For an easy comparison between different calculations and codes, a practical choice is to represent all band structures along the paths defined in the paper by W. Setyawan and S. Curtarolo\cite{Bands}, if these were calculated.
Other properties of general interest are the density of electronic states and the effective masses. 
For electronic band structures, density of states (DOS), and related quantities, the energy zero can be conveniently set to the highest occupied Kohn-Sham level.\\
Calculations of harmonic vibrations in bulk materials (phonons) store all relevant information either in real space (force-constant matrix) or in reciprocal space (appropriate set of dynamical matrices), whereby it is important to note that, for non-polar materials, these two quantities are unambiguously related to each other via Fourier transformations. For polar materials, this data might be augmented by the dielectric tensor and the Born effective charges, which affect the long-wavelength vibrations in such materials.
For calculations that contain them, force-constant matrices or dynamical matrices can be efficiently stored and they allow to compute other vibrational properties with negligible computational effort. For instance, vibrational band structures along the exact same paths used for the electronic band structures, but also densities of vibrational states and thermodynamic properties such as specific heats can be easily derived. Obviously, an identifier of the original calculations used to obtain the vibrational properties should be stored along with the discussed vibrational properties. This ensures that the employed computational and physical settings can be retrieved, if needed.

\subsection{Compact representation of scalar fields: density, wavefunction, xc potentials, etc.} 
The comparison of scalar fields across methodologies and codes requires to translate the internal, code and basis set specific representation of these fields into a common format. For such a representation, an all-electron formalism is desirable, since it allows to evaluate additional properties such as electric field gradients and NMR shifts.\\
Different codes use different basis functions which can be divided into two subclasses: localized functions and periodic functions. Popular choices for localized functions are Gaussian-type orbitals and numeric atomic orbitals (NAOs), while for periodic functions plane-waves or augmented plane waves are often used. Plane-wave basis sets are usually combined with pseudopotentials. Approximate all-electron wavefunctions can be straightforwardly restored from such pseudopotential calculations\cite{psptoae}.\\
To represent wavefunctions in a code-independent format, conversion to a universal basis set can be used. The conversion from the original basis $\phi_\alpha$ to the universal basis $\eta_\beta$ is performed by solving:
\begin{equation}
  \sum_\gamma S_{\beta\gamma}\tilde{C}_\lambda^i = \sum_\alpha \langle \eta_\beta |\phi_\alpha \rangle C_\alpha^i \, ,
\end{equation}
where $\tilde{C}_\lambda^i$ and $C_\alpha^i$ are the coefficients of the expansion in the new and original basis, respectively, and ${\bf S}$ is the overlap matrix for the universal basis functions. Additional constraints can be taken into account when minimizing differences between the original and the universal representation, such as strict orthonormalization of the transformed wavefunctions.\\
The following two choices are most promising candidates for the universal basis:
\begin{itemize}[noitemsep,nolistsep]
\item Gaussian basis functions. Online libraries of Gaussian basis sets are available (e.g., \\ \texttt{https://bse.pnl.gov/bse/portal}).
\item NAOs constructed once and for all for each species. A hierarchical construction (similar to Gaussian basis sets) of highly optimized NAO basis sets of various sizes is implemented in some codes, such as DMol$^3$ and FHI-aims.
\end{itemize}
Which specific all-electron basis set is best suited for this purpose will be evaluated in detail in an upcoming publication\cite{wavefunc}.

In addition to the code-independent representation, a common storage format that allows for a quick restoration of scalar fields in the native, code-specific representation is useful, for example for restarts from previous calculations. Such representation can be very compact because often only part of the information needs to be stored. The missing part can then be quickly obtained on the fly by running the code. For example, storing only plane-wave coefficients is sufficient for VASP to restore localized functions describing the wavefunctions near a nucleus, whereas an explicit storage of these localized functions would be very demanding. Since the concept of representing scalar fields in a basis-set expansion is common for all electronic-structure codes, defining a common, code-independent file format for storing the restart information for different codes is possible. Namely, one can store information identifying the functional form of a basis function (plane wave, contracted Gaussian, etc.), and the corresponding expansion coefficient. Obviously, such a representation cannot be used for code-independent data analysis, but it is beneficial from the practical point of view.

\subsection{Quantities related to excited-state calculations.} 
Advanced many-body perturbation theory (MBPT) calculations ($GW$, Bethe-Salpeter equation, etc.) currently output only few properties (spectra, self-energies, etc.) that need to be parsed and stored.  To facilitate the analysis of this kind of calculations, it is essential to develop and store a detailed classification of all approximations used in the MBPT calculation in the metadata, given that many different numerical formalisms are implemented in different MBPT codes.\\
The $GW$ approach is nowadays considered a routine approach for computing quasi-particle band structures. However, what is generically called $GW$ comprises a lot of different approximations (besides those of the underlying ground state calculations). Presently, the situation is less transparent than for ground-state. For this reason, it is necessary to store the following information:
\begin{itemize}[noitemsep,nolistsep]
\item starting point (xc functional of the underlying DFT calculation), 
\item whether the calculation has been carried out in a perturbative manner or self-consistently (in fact, several types of self-consistency have been developed), 
\item further approximations, like plasmon-pole models,
\item auxiliary basis sets used for non-local operators,
\item numerical approximations, such as size of reciprocal space meshes, basis set size, frequency/time grid settings, etc.,
\item whether the sum over unoccupied states is avoided / truncated / approximated,
\item whether involved quantities are represented in real or reciprocal space, 
\item whether the Coulomb potential is truncated.
\end{itemize}
Obviously, all these approximations must be labelled by appropriate metadata (see also Section 2.1). 
$GW$-related quantities of interest to be stored are:
\begin{itemize}[noitemsep,nolistsep]
 \item matrix elements of the exchange and correlation contributions to the self-energies evaluated at the Kohn-Sham states in case of $G_0W_0$. For self-consistent $GW$ calculations, the matrix elements are evaluated with quasi-particle states.
 \item matrix elements of the xc potential,
 \item quasi-particle energies,
 \item spectral functions (if calculated).
\end{itemize}
For optical spectra determined by TDDFT or the solution of the Bethe-Salpeter equation the situation is similar to what was described above. The quantities of interest are:
\begin{itemize}[noitemsep,nolistsep]
 \item excitation spectra, energies, and oscillator strengths,
 \item exciton binding energies (if available),
 \item spectra obtained by the independent-particle approximation and/or KS response function.
\end{itemize}

\subsection{Establishing error bars, uncertainties, and confidence levels.} 
Quantifying the errors and uncertainties of the data included in computational materials' databases is an essential step to make this data useful.  Challenges in this field arise, since the errors are code, material, and even property specific. Also, the dependence of different errors on each other needs to be taken into account. A first step in this direction is to establish unique identifiers for structures through ``similarity recognition''. In this way, group of calculations performed on atomic structures that are not identical but obtained by small distortions of the coordinates and/or the unit cell can be automatically recognized as similar and an error analysis can be performed. Furthermore, a systematic investigation of numerical errors is required across codes for both simple and complex properties, which also requires a clear definition of errors/deviances, e.g., for continuous functions. With respect to errors arising from the use of approximated xc-functionals, (more) test sets are required as a reliable, high-level reference. 
The use of an experimental benchmark is avoided, for the moment, as temperature and pressure conditions, intrinsic defects and impurities, surfaces, or
dislocations can make it difficult to obtain unambiguous (test) sets of reference values from experiments.\\
For a given atomic configuration, there are different error bars corresponding to the different approximations: 
\begin{itemize}[noitemsep,nolistsep]
 \item basis set, 
 \item choice of pseudopotential (if employed),
 \item grids and other numerical approximations, 
 \item $k$-mesh,
 \item treatment of relativity,
  \item xc functional.
\end{itemize}
Thus, every calculated result stored in a code-independent format should be connected using the method-related metadata with six numbers that refer to the mentioned (mostly code-specific) approximations.
Obviously, energies and energy differences will be associated with different error bars, just to mention one example. In general, the importance of these error bars depends on the material's property of interest. \\
The mentioned error-bar contributions may be evaluated from any dataset that contains results corresponding to the same material, but using different approximations and/or different codes. However, it will also be necessary to evaluate relevant quantities at different levels of approximations systematically. In fact, this will also yield a ``test set for materials'' which is a well-known and most useful concept in quantum chemistry but largely absent, so far, in materials science. The build-up of a ``test set for materials science and engineering'' has been initiated. 
Some such studies are being performed by the groups of G. Kresse in Vienna, and others by the NOMAD team \cite{errors,MSE}.\\
In addition to the numerical errors discussed above, it is essential to develop means to assess the possible error coming from the actual implementation (coding) of the various atomic-scale calculations. This requires the establishment of high-quality benchmark calculations for various materials and properties. For ground-state calculations, the work by Cottenier and co-workers \cite{Cottenier, Cottenier2} for bulk materials represents a first and important step in this direction. The next step, which considers low symmetry situations, e.g. defects and surfaces is just initiated. For excited-state codes the GW100 paper\cite{GW100} comparing TURBOMOL, FHI-aims, and BerkleyGW for quasi-particle energies of molecules, represents a first step towards this goal. 

\section{Conclusions}
We have presented challenges and practical strategies for achieving a common format for the representation of computational materials science data. 
The strategy goes through the definition of a metadata infrastructure that can be used for writing a conversion layer from existing codes to a code-independent format (the NOMAD Laboratory CoE), as well as to provide a library for the direct output of present and future electronic structure codes into a standard format (the ESCDF initiative). For several crucial topics, like the common energy zero-level, electronic and vibrational properties, the scalar-field representation, and excited states calculations, we present practical choices for achieving our goal of a code-independent representation.
We also present the challenge of establishing reliable error bars and uncertainties for each calculation stored in a database, in terms of the adopted numerical settings.

\section{Acknowledgements}
This project has received funding from the European Union's Horizon 2020 research and innovation program under grant agreement No 676580, The NOMAD Laboratory, a European Center of Excellence, and the BBDC (contract 01IS14013E).\\
We thank James Kermode and Saulius Gra\v{z}ulis for their contribution to the discussion on the metadata, and Pasquale Pavone for precious suggestions on the metadata structure and names.\\
We thank Patrick Rinke for carefully reading the manuscript.\\
We thank Claudia Draxl and Kristian Thygesen for their contribution to the discussions on the necessary information to be stored for excited-state calculations and on the error bars and uncertainties.\\
We gratefully acknowledge Damien Caliste, Fabiano Corsetti, Hubert Ebert, Jan Minar, Yann Pouillon, Thomas Ruh, David Strubbe, and Marc Torrent for their contributions to the ESCDF specifications.\\
We acknowledge inspiring discussions with Georg Kresse, Peter Blaha, Xavier Gonze, Bernard Delley, and J\"{o}rg Hutter on the energy-zero definition and scalar-field representation.\\
We thank Ole Andersen, Evert Jan Baerends, Peter Blaha, Lambert Colin, Bernard Delley, Thierry Deutsch, Claudia Draxl, John Kay Dewhurst, Roberto Dovesi, Paolo Giannozzi, Mike Gillan, Xavier Gonze, Michael Frisch, Martin Head-Gordon, Juerg Hutter, Klaus Koepernik, Georg Kresse, Roland Lindh, Hans Lischka, Andrea Marini, Todd Martinez, Jens J\o{}rgen Mortensen, Frank Neese, Richard Needs, Taisuke Ozaki, Mike Payne, Angel Rubio, Trond Saue, Chris Skylaris, Jose Soler, John Stanton, James Stewart, Marat Valiev for checking the information provided in Table 1 and for useful suggestions.

\bibliographystyle{unsrt}


\end{document}